\documentclass[10pt, letterpaper]{article}
\usepackage[margin=1in]{geometry}
\usepackage{graphicx}
\usepackage{amsmath}
\usepackage{cite}
\usepackage{authblk}
\usepackage{url}
\usepackage{kantlipsum,setspace}
\usepackage{caption}
\usepackage{siunitx,booktabs}
\newcommand{\mlc}[1]{\multicolumn{1}{c}{#1}}

\pdfoutput=1
\title{Cholera forecast for Dhaka, Bangladesh, with the 2016 El Ni\~no}

\author[1]{Pamela P. Martinez}
\author[2]{Robert C. Reiner Jr.} 
\author[3]{Manojit Roy} 
\author[4]{Benjamin A. Cash}
\author[5]{Md. Yunus} 
\author[5]{A.S.G. Faruque} 
\author[5]{Sayeeda Huq} 
\author[3,6]{Aaron A. King } 
\author[1,7]{Mercedes Pascual\thanks{pascualmm@uchicago.edu}}
\date{\today}
\affil[1]{Department of Ecology and Evolution, University of Chicago, Chicago, IL 60637}
\affil[2]{Department of Epidemiology and Biostatistics, Indiana University Bloomington School of Public Health, Bloomington, IN 47405}
\affil[3]{Department of Ecology and Evolutionary Biology, University of Michigan, Ann Arbor, MI 48109}
\affil[4]{Center for Ocean-Land-Atmosphere Studies, George Mason University, Fairfax, VA, 22030}
\affil[5]{International Centre for Diarrheal Disease Research, Dhaka 1000, Bangladesh}
\affil[6]{Department of Mathematics, University of Michigan, Ann Arbor, MI 48109}
\affil[7]{Santa Fe Institute, Santa Fe, NM 87501}

\begin{document}

\maketitle

\section*{Abstract}
A substantial body of work supports a teleconnection between the El Ni\~no Southern Oscillation (ENSO) and cholera incidence in Bangladesh. In particular, high positive anomalies during the winter (December-January-February) in Sea Surface Temperatures (SST) in the Tropical Pacific have been shown to exacerbate the seasonal outbreak of cholera following the monsoons from August to November, and climate studies have indicated a role of regional precipitation over Bangladesh in mediating this long-distance effect. Thus, the current strong El Ni\~no has the potential to significantly increase cholera risk this year in Dhaka, Bangladesh, where the last five years have experienced low seasons of the disease. To examine this possibility and produce a forecast for the city, we considered two models for the transmission dynamics of cholera: a statistical spatio-temporal model previously developed for the disease in this region, and a process-based temporal model presented here that includes the effect of SST anomalies in the force of infection and is fitted to extensive cholera surveillance record between 1995 and 2010. Prediction accuracy was evaluated with `out-of-fit' data from the same surveillance efforts (post 2008 and 2010 for the two models respectively), by comparing the total number of cholera cases observed for the season to those predicted by model simulations eight to twelve months ahead, starting in January each year. Encouraged by accurate forecasts for the low risk of cholera for this period, we then generated a prediction for this coming season. An increase above the third quantile (75\%) in cholera cases is expected for the period of August - December 2016 with 92\% and 87\% probability respectively for the two models. This alert warrants the preparedness of the public health system. We discuss the possible limitations of our approach, including variations in the impact of El Ni\~no events, and the importance of this large, warm event for further informing an early-warning system for cholera in Dhaka.

\newpage
\section*{Introduction}
\begin{figure*}[b!]
\begin{center}
\centerline{\includegraphics[width=.875\textwidth]{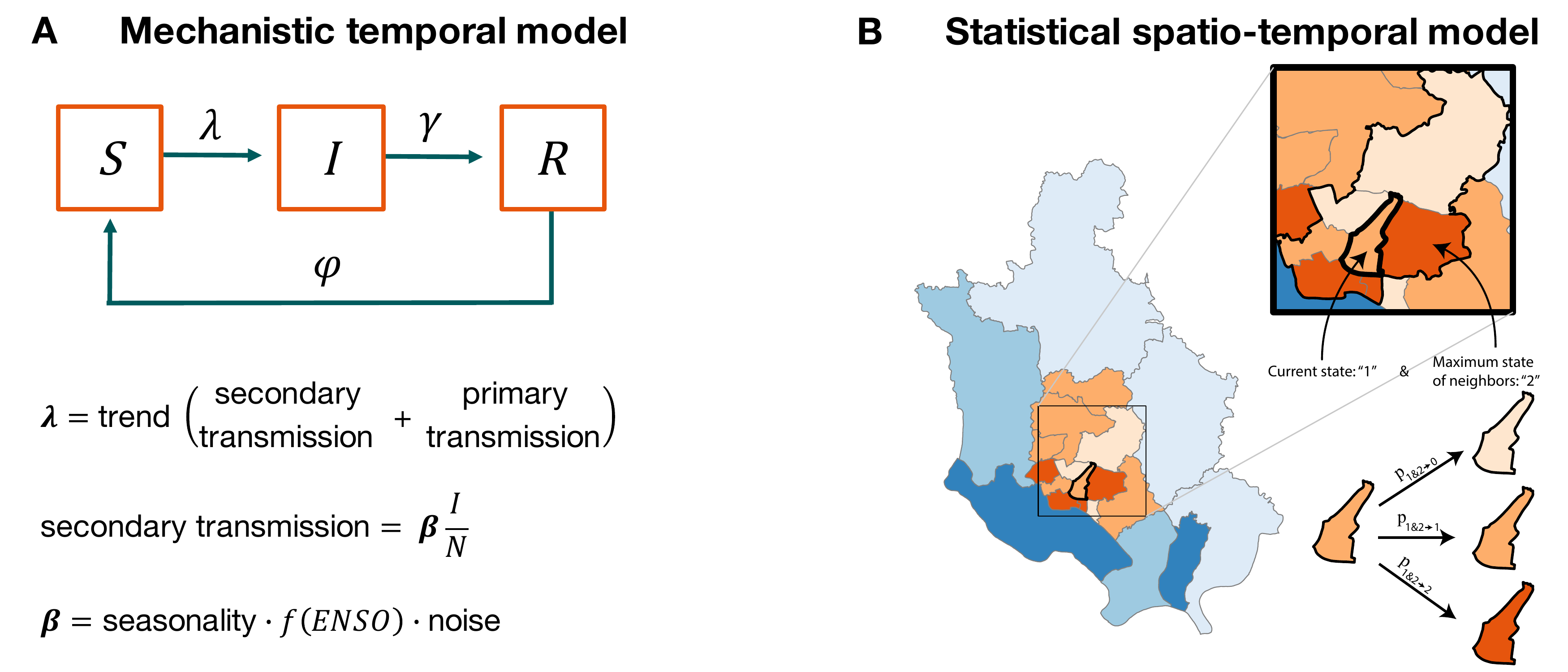}}
\caption{\textbf{Schematic representation of the models. A)} Mechanistic temporal model. The population is divided into three classes, for the susceptible ($S$), infected ($I$) and recovered ($R$) individuals. The arrows denote rates of flow among these classes. The force of infection ($\lambda$) includes three components: a long-term trend, secondary transmission that depends on the levels of infection in the population, and primary transmission at a constant rate from an environmental reservoir of the pathogen. The transmission coefficient (or rate) $\beta$ in secondary transmission incorporates seasonality, interannual variation as a function of the ENSO index, and environmental noise. \textbf{B)} Statistical spatio-temporal model. Districts of the city known as `thanas' are grouped into two main regions (depicted in orange for the core and in blue for the periphery). Thanas within the same group follow the same dynamical rules in terms of transitions between cholera levels or states from one month to the next. Three states are considered and used to discretize the case data: no cholera (0), low cholera (1), or high cholera (2) as indicated by the different color intensity. The probability of transition between states from one month to the next depends on the season, the maximum state of neighboring districts, and the climate covariate (ENSO). For details on the models, see Methods and Supplementary Information.
}\label{figure1}
\end{center}
\end{figure*}

Climate variability, specifically the El Ni\~no Southern Oscillation (ENSO), has been shown to influence the year-to-year variation of seasonal outbreaks of cholera in Bangladesh \cite{colwell1996, pascual2000, bouma2001, rodo2002, koelle2005, pascual2008}. In particular, this interannual variation of the disease has been shown to be positively associated with Sea Surface Temperatures in the Tropical Pacific in the region that warms during El Ni\~nos \cite{pascual2000}. Regional rainfall was identified as one of the local drivers that mediates this long-range effect of ENSO \cite{cash2008}, with evidence for an influence of high monsoon rains and river discharge on disease levels in Bangladesh (e.g. \cite{matsuda2008, hashizume2008, akanda2009, hashizume2009, akanda2011, nasr2015, jutla2015}).

Both statistical analyses and process-based mathematical models for the population dynamics of the disease have been applied to the role of ENSO in rural areas of Bangladesh (e.g. \cite{pascual2000, koelle2005}). For the city of Dhaka itself, the capital of Bangladesh, application of mechanistic models incorporating epidemiological processes is still lacking, although a more phenomenological approach has been developed to analyze the spatio-temporal transmission dynamics at the level of `thanas' or districts within the city \cite{reiner2012}. The statistical model of Reiner \textit{et al}. (2012) specifically identified the existence of two distinct spatial regions within Dhaka based on the dynamics of the disease: the truly urban core of the city and its more rural periphery, with the former exhibiting much higher disease incidence and a stronger response to ENSO. This distinction between the core and periphery of the city was also found to be relevant for rotavirus, suggesting that the causal links between climate and disease are a more general feature of diarrheal diseases in this region, independent of their particular transmission pathways \cite{martinez2016}. A similar conclusion is supported by comparisons between cholera and shigellosis regarding their association with flooding and SST in the Pacific \cite{cash2014}.

Recently, new statistical approaches for inferring model parameters based on time series data have allowed flexible representations of mechanisms that more closely represent epidemiology in models that also incorporate both measurement and process noise \cite{ionides2006, king2016}. Although these process-based models have been applied to glean insight into epidemiological processes in a number of infectious diseases from retrospective data, they have been used only in a few exceptions to generate and evaluate forecasts (e.g. \cite{roy2015, king2015}). Motivated by the large El Ni\~no of 2015-2016, we examine here the ability of a process-based (mechanistic) model to predict cholera incidence in Dhaka by combining epidemiology and climate variability (Figure 1A). We also compare the results to those obtained with the previously published statistical model for cholera in the city (Figure 1B). The value of the SST anomaly observed in the region of the Pacific known as Nino3.4 for this January (2.60) is comparable in magnitude to the one observed for the large El Ni\~no event of 1998 (2.56), when Dhaka suffered the largest cholera outbreak in the last 20 years and one of the worst floods in its history affecting more than 50\% of the city's area \cite{faisal2003}. We focus on cholera reported cases from the core of the city for the process-based model, based on the heightened sensitivity to ENSO in this region, the low number of cases in the periphery and its low contribution to the force of infection in the core \cite{reiner2012, perez2016}. The monthly surveillance data are divided into a training set from January 1995 to December 2010, used to fit the mechanistic model, and an `out-of-fit' set from January 2011 to December 2015, used to evaluate the predictability of the model. For the statistical model, the out-if-fit set starts in 2009, based on the published fit \cite{reiner2012}. We specifically ask whether these two models can be used as an early-warning system, able to forecast cholera outbreaks following the monsoon season, based on observed January Ni\~no3.4 values and on previous cholera cases. We then generate a forecast for the incoming transmission season in August-December 2016 and discuss implications of the findings and possible limitations of the approach given the event variability of El Ni\~no and of its connections to the regional precipitation over Bangladesh. We discuss what will be learnt later this year when the actual observations for the upcoming cholera season become available, whether the predictions fail or succeed. At the moment, careful monitoring of the development of the monsoon season is warranted in informing cholera predictions at a shorter lead time.

\section*{Results}
Normalized incidence data exhibit strong interannual variability (Figure 2). The seasonality in cholera incidence is bimodal with two peaks per year: in spring and in autumn, preceding and following the monsoons respectively. For the mechanistic model, the observed and the simulated data exhibit yearly peaks and troughs that fall in the same months for most years, capturing the bi-modal seasonal patterns fairly well (Figure 2A). Such agreement is encouraging given that these simulations are not one time-step ahead predictions but instead, 15-yr long trajectories starting from estimated initial conditions. It demonstrates the hindcast performance of the model and suggests its potential for forecasting. 

\begin{figure*}[t!]
\begin{center}
\centerline{\includegraphics[width=.98\textwidth]{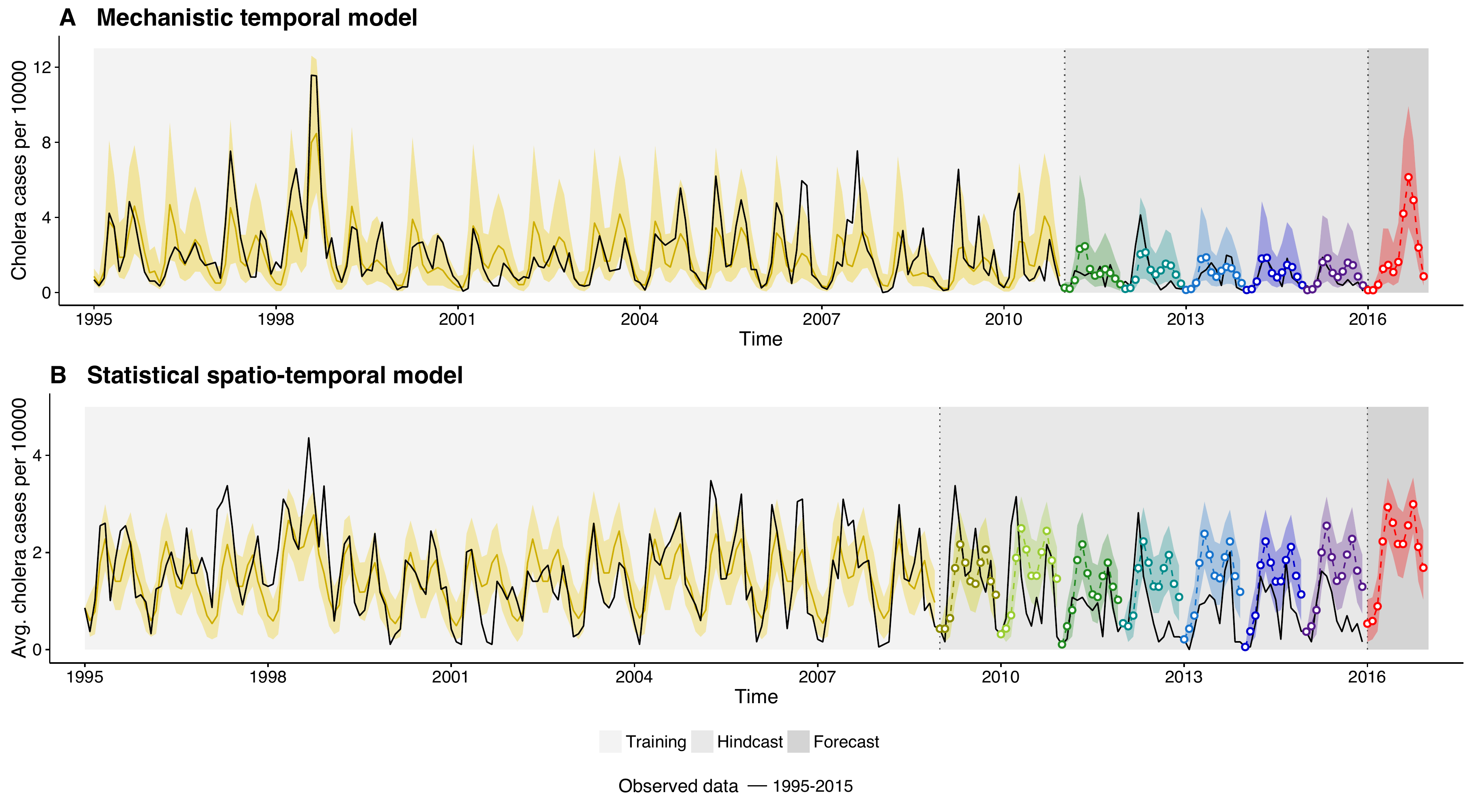}}
\caption{\textbf{Comparison of simulated and predicted monthly cases with those reported for Dhaka, Bangladesh}. The observed cases per 10000 individuals are shown in black for the core of the city (A) and for the average of all thanas (for core and periphery, (B)). The median of 1000 simulations is shown in dark yellow, with the 10-90\% confidence intervals (C.I.) in the shaded lighter color. Simulations of the predicted out-of-fit data are shown in different colors starting in 2011 in (A) and in 2009 in (B), with the median for each month in open circles, and their respective C.I. envelope in a shaded lighter color. Vertical dotted lines and corresponding background shading indicate three different kinds of model simulations and `predictions': for the initial period (white background), the model is simulated from 1995 forward. Thus, the comparison to data is not based on the typical next-step (next-month) prediction for which it is somewhat trivial for most models to capture the fitted data. These simulations span more than a decade. The second period (light gray background) shows predictions for out-of-fit years, with simulations starting in January and spanning the whole year. We refer to these predictions that cover windows of time in the past, as hindcasts. Finally, the last period is for the current year and constitutes a true forecast (for the upcoming fall season).
}\label{figure2}
\end{center}
\end{figure*}

The predicted yearly data for the period 2011-2016 were generated by updating the initial states of the epidemiological variables for January each year, as well as the estimated values of the parameters for each year starting in 2011. Model predictability was evaluated for the years between 2011 and 2015. Because the model is stochastic (incorporating noise in the dynamics), we generated both the median value and the 10-90\% confidence intervals of 1000 predictions for each month. For the most part, the median prediction appears close to the data, with the observed cases falling within the uncertainty of the confidence intervals, suggesting that cholera incidence is forecasted accurately, including the decline observed in the data until 2015 and the low levels of incidence characteristic of the past 5 years. We note a slight over-prediction of some of the fall seasons (2012 and 2015) which exhibited almost no cases those years in a somewhat uncharacteristic seasonal pattern. Similarly, simulations of the statistical model show that the predicted cases are very consistent with the observed data within the training set as already shown in Reiner \textit{et al}. (2012) for data up to 2009 (Figure 2B). Here, we can evaluate the ability of this model to predict the new data not used to fit the model, which has become available since that study. For the most part, the predictions produce the observed lower levels of cholera, with an overprediction of the same fall peaks for 2012 and 2015. Here the overprediction is larger possibly because the model does not incorporate a long-term trend.

To more formally evaluate predictability, we considered the probability of exceeding a threshold number of cases for the whole fall season (Aug - Dec), with the threshold based on the distribution of outbreak size in the data. We are interested here in evaluating the ability of the model to predict the occurrence vs. the lack of `large' outbreaks, given their relevance to public health and in light of the high El Ni\~no index reported for January 2016. In Table 1, the probability of cholera incidence exceeding the 50th, 75th and 95th percentiles for 2011-2015 are reported, with these different thresholds representing outbreaks of increasingly higher magnitude. Results from the mechanistic model indicate the absence of a large outbreak, consistent with the observed data for that period, with none of the years exhibiting a probability greater than 50\% of exceeding the thresholds (Table 1). In other words, the model would have accurately predicted the low risk of cholera in the past five years. Likewise, the statistical model shows consistent results for the most part, with a probability higher than 50\% of being above the median for only one of the years (2015). Otherwise for the 75\% and 95\% thresholds, the probabilities of exceeding them are consistently low. For the coming fall season of 2016, the same forecast analysis reveals probabilities of 87\% and 92\% for the mechanistic and statistical models respectively, of a large outbreak (defined as surpassing the 75th quantile of outbreak size). For an extreme outbreak (defined as surpassing the 95th quantile, and therefore comparable to the cholera event of 1998), these probabilities are 50\% and 27\% for the two models. We note that the latter model has a tendency to underpredict the actual size of large outbreaks, evident for 1998. As explained in Reiner \textit{et al}. (2012), this follows from the classification into three discrete levels which tends to place all thanas with a high number of cases in the same category. This effectively reduces the tail of the distribution of monthly cases. We return to this consideration in the Discussion.

\begin{table}[b!]
\def\arraystretch{0.9}
\addtolength{\tabcolsep}{2.5pt}
\centering
\caption{\textbf{Hindcasts for the indicated years and forecast for 2016, for the post-monsoon (Aug-Dec) season of cholera}. The distribution of observed cases for this same post-monsoon period for the training data used to fit the models was used to estimate the values of the 50th (the median), 75th and 95th quantiles. These values are used as thresholds to define outbreaks of increasing size: a season that exceeds the median is considered anomalous, one that exceeds the 75\% level is considered a large outbreak, and one that exceeds the 95\% level, an extreme outbreak. The average observed cases for each year are shown next with an indication of whether they exceed the threshold (yes, ``outbreak'') or not (no, ``no outbreak''). The proportion of 1000 simulations that fall above each threshold level is reported as a probability, and a probability $> 50\%$ is interpreted as a prediction of an outbreak, specified in the last column of each model.}
 \begin{tabular}{cccccccc}
\toprule[\heavyrulewidth]
&  & \multicolumn{3}{c}{\textbf{Mechanistic temporal model}} 
& \multicolumn{3}{c}{\textbf{Statistical spatio-temporal model}}\\[0.1cm]
\cmidrule(rl){3-5} \cmidrule(rl){6-8}
\cmidrule(rl){3-5} \cmidrule(rl){6-8}
\mlc{Year} & \mlc{Quantile}  
& \mlc{Observed} & \mlc{Probability} & \mlc{Prediction}
& \mlc{Observed} & \mlc{Probability} & \mlc{Prediction}\\
\midrule[\heavyrulewidth]
&  50  & no & 14.3 & no& no& 3.4 & no\\
2011&  75  & no& 1.3 & no& no& 0.5 & no\\
&  95 & no& 0.1 & no& no& 0.0 & no\\
\midrule
&  50  & no& 17.6 & no& no& 14.5 & no\\
2012&  75  & no& 2.0 & no& no& 4.4 & no\\
&  95 & no& 0.1 & no& no& 0.0 & no\\
\midrule
&  50  & no& 19.0 & no& no& 45.3 & no\\
2013&  75  & no& 1.5 & no& no& 23.0 & no\\
&  95 & no& 0.0 & no& no& 0.0 & no\\
\midrule
&  50  & no& 18.7 & no& no& 35.5 & no\\
2014&  75  & no& 1.6 & no& no& 13.0 & no\\
&  95 & no& 0.0 & no& no& 0.0 & no\\
\midrule
&  50  & no& 18.3 & no& no& 55.6  & yes\\
2015&  75  & no& 1.8 & no& no& 29.0 & no\\
&  95 & no& 0.1 & no& no& 0.0 & no\\
\midrule[\heavyrulewidth]
&  50  & -- & 99.0 & yes & --  & 99.2 & yes\\
\textbf{2016}&  75  & -- & 86.5 & yes & -- & 97.2 & yes\\
&  95 & --  & 50.1 & yes &  -- & 26.5 & no\\
\midrule[\heavyrulewidth]
\end{tabular}
\end{table}

We also consider the evolution of the monsoon over Bangladesh up to this time in June, based on the Climate Prediction Center Unified (CPC\_UNI) global daily precipitation data set \cite{xie2007, chen2008}, which covers the period 1979-present and is updated daily in real time. We compare the evolution of the 2016 monsoon with previous high and low flooding years, all following El Ni\~no events. Five high flooding (1987, 1988, 1998, 2004, 2007) and five low flooding (1983, 1992, 1995, 2003, 2005) years were selected, and the average difference in June rainfall relative to the 1979-2015 mean is shown in Figure 3. We find that for the high flooding years, June rainfall was higher than normal across much of Bangladesh (Figure 3A), while in the low flooding years, June rainfall was below normal across the entire region (Figure 3B). Signals from the current year are more mixed, with above normal rainfall across the Northwest India and northern Bangladesh and below normal rainfall across the center of the country (Figure 3C). At the time of writing, the Bangladesh Flood Warning and Forecast Center (http://www.ffwc.gov.bd/) reports below normal rainfall across most of the major river basins.

\section*{Discussion} 
\begin{figure*}[b!]
\begin{center}
\centerline{\includegraphics[width=.95\textwidth]{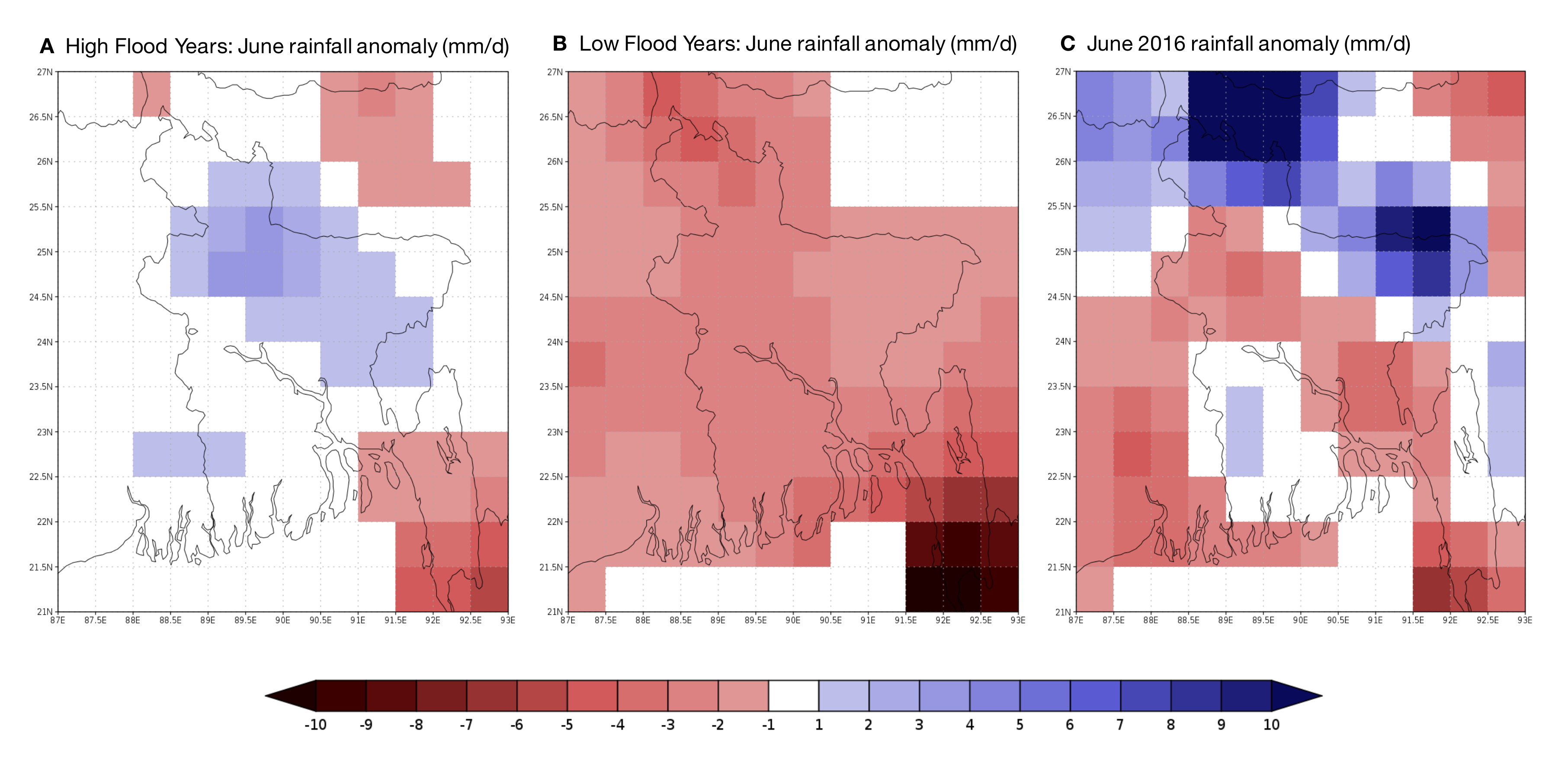}}
\caption{\textbf{Deviation from climatological June rainfall for A) five high flooding years, B) five low flooding years, and C) 2016}. Units are (mm/d).}\label{figure3}
\end{center}
\end{figure*}

The El Ni\~no Southern Oscillation (ENSO) acts as the main driver of interannual climate variability worldwide. As such, the associated anomalies in Sea Surface Temperatures in the Tropical Pacific provide a basis for predicting the interannual variability of a number of phenomena around the globe, including that of climate-sensitive infectious diseases, such as water-borne and vector-borne infections. Here, we implemented a mechanistic transmission model that incorporates both epidemiological processes and the effect of ENSO, to forecast cholera risk in the city of Dhaka for the upcoming fall season. The predictability of this model was validated with retrospective data, and our main results indicate an 87\% probability to observe a large outbreak of cholera during Aug-December 2016, comparable to that of the previous large El Ni\~no in 1998. Likewise, the previously published statistical model for cholera in Dhaka indicates a 92\% probability of a large outbreak this coming fall. These results provide a call for a heightened alert and preparedness of the public health system in the city, concurrent with the monitoring of the evolution of the monsoon season in Bangladesh. 

The predictive ability of the models was evaluated based on the incidence of the last five years. Given the uniformly low levels of cholera during that period, this assessment concerns only true or false negatives, that is the ability of the model to predict the lack, and not the occurrence, of an outbreak. Although we correctly predict the lack of an outbreak and the overall low risk of cholera for that period, we overpredict the actual number of monthly cases during the fall season for two of the years (2012 and 2015) whose fall peaks were almost completely absent. This pattern may result from the stochasticity inherent to low levels of the disease, and in part be accounted for the long-term trend in combination with the lack of warm anomalies in ENSO conditions. The statistical model overpredicts these recent low seasons further, possibly because it does not incorporate such a trend. We note that in general, this model also tends to underpredict the large events (like 1998) because of the transformation from actual cases to three discrete levels which reduces the tail of the distribution of cases \cite{reiner2012}. A detailed analysis of this property of the model showed that it can be corrected by making a probability higher than 25\% (rather than 50\%) an indication of outbreak risk (for the 75\% threshold). We note that this correction does not modify the outcome for 2016, and that the probability for the 75\% quantile is 97\%, well above 25\%, a further indication that the model predicts a high risk of a large outbreak. 

Our predictions are based on the magnitude of the current El Ni\~no condition, which is similar to that observed for 1998 when Dhaka experienced the largest cholera outbreak in the last 20 years. Given that the interannual variability associated with ENSO is superimposed on the on-going climate change, the effect of El Ni\~no on climate-sensitive diseases could be aggravated. We acknowledge, however, that variation among El Ni\~no events also calls for caution regarding predictions that involve long-distance connections, such as ours, since such an event is not fully characterized by the magnitude of the SST anomaly alone, and ultimately, the regional changes in the climate variability of Bangladesh are what matters to the population dynamics of the disease. The ENSO index permits however the generation of forecasts with a relatively long lead time, of 8 to 12 months, not possible for any local environmental factor. This early warning can then be followed by carefully monitoring of the monsoon especially in July and August. 

Given that some strong El Ni\~no events have resulted in low rainfall and low flooding in Bangladesh in the past, and the somewhat anomalous impact of the 2015-2016 event on rainfall in other regions (such as California), it is important to recall that no ENSO index can fully capture the impact of a given event. Significant difference in impact over Bangladesh from events characterized by similar index values have occurred in the past. These observations are a reason for caution, indicating the possibility of a false positive forecast for cholera if rainfall were to remain below average for the whole season.

We note however that the value of producing the forecast goes beyond its actual success since we can learn from comparing it to the actual incidence of cholera, once the data become available, whether or not a large outbreak occurs. Specifically, the lack of a large cholera outbreak could follow from three possible reasons: First, the current El Ni\~no could differ from previous ones such as 1998 in its effects over Bangladesh. In this case, future research should seek further features of the evolution of such an event that can be useful to distinguish effects on cholera in the region, although this pertains to the challenging connections between temperatures in the Pacific and the monsoons. Still, the discrepancy of model predictions and observations would be an indication of the importance of concatenating an early warning with a long lead time, to a shorter prediction based on the local environment. Second, a discrepancy with our forecast could result from increased and effective intervention. In this case, it would be valuable to document the level of intervention in comparison for example to that of 1998. Model outputs, besides their potential contribution to preparedness, can serve as a point of reference to evaluate the success of intervention \cite{roy2015}. Finally, the low incidence observed in the last 5 years, especially for the fall peak, could be a consequence of changes in the environmental and/or socio-economic conditions in the city, which could alter the seasonality and overall risk of the disease. In this case, even in the presence of anomalous rains, one would not observe as strong a response to climate forcing as those in the past. A related intriguing possibility is that the prolonged hiatus in large El Ni\~nos has contributed to the sustained window of many successive years with low incidence, which in turn have diminished the reservoir of the bacterium \textit{Vibrio cholerae} in its pathogenic form. In this case, the system may be poised at environmental conditions that make it unable to respond as strongly as in the past to a large climate event. Observations of how both the disease and the monsoons evolve this year can tell us which of these situations apply. 

In brief, we provide the forecasts for the upcoming season as a warning, based on the past accuracy of the models and on the lead time afforded by observations of SST in the Tropical Pacific. Public health preparedness is called for, to be complemented by observation of the monsoon in Bangladesh in July-August, and later still, by observations of the cholera cases themselves as the transmission season develops. With the more sophisticated ability of the modeling community to fit transmission models to retrospective records, it is now of interest to take up the challenge of forecasting and learn from `real time' implementations. 

\section*{Methods}
The original cholera data consist of daily cases from 1995 to 2015 obtained from the ongoing surveillance program by the International Center for Diarrheal Disease Research, Bangladesh (icddr,b), in which a systematic subsample of all patients visiting the hospital, which serves as the main treatment center for the greater Dhaka city area, is tested for cholera. Reported cases were added by month, per `thana' or administrative subdivision. For each thana, the population was computed by interpolating the three decadal censuses beginning in year 1991, 2001 and 2011. Cases were then aggregated over the thanas comprising two regions, core and the periphery of the city, according to the partition proposed by Reiner \textit{et al}. (2012). 

The process-based model follows the temporal changes in the number of cases in the core of the city. It consists of a Susceptible-Infected-Recovered-Susceptible (SIRS) compartmental formulation in which the population is subdivided into the following classes: $S$ for susceptible individuals, $I$ for infected and infectious individuals, and $R$ for recovered individuals who have acquired immunity to the disease. Acquired immunity is temporary and wanes at rate $\phi$ as individuals return to the S class. The recovery rate ($\gamma$) of infected individuals and other parameters are estimated using a recently developed likelihood-based inference method (see Supplementary Information for model details). The rate of transmission per susceptible individual (or force of infection $\lambda$) contains three components: a long-term trend, a representation of secondary transmission (depending on the level of infected individuals in the population), and primary transmission (at a constant rate from environmental reservoir). Secondary transmission depends on the proportion of the population that is infected ($I/N$), with the coefficient or transmission rate $\beta$ including a seasonal component, the interannual effect of ENSO, and environmental noise (figure 1, figure S1). The model further incorporates measurement error with a reporting rate that is also estimated and accounts for under-reporting. For a full description of the model, see Supplementary Information. 

The statistical model follows the spatio-temporal transmission dynamics of the disease at the spatial resolution of thanas or districts in the city. It categorizes the monthly thana-level case data into three states: `no cholera'; `low cholera'; and `high cholera' as described in Reiner \textit{et al}. (2012). The model estimates the month-to-month transitions of each thana as a set of three transition probabilities (conditioned on the value of the current state only). These probabilities are functions of temporal and spatial covariates as well as of the maximum state of the neighboring thanas (Figure 1). The parameter values used for these predictions were fitted to data from 1995-2008 in Reiner \textit{et al}. (2012). For complete details on the model formulation and fitting, including the categorization criterion, see Reiner \textit{et al}. (2012).

\section*{Acknowledgements}
The case data used in this paper were collected with the support of icddr,b and its donors who provide unrestricted support to icddr,b for its operation and research.  Current donors providing unrestricted support include the Government of the People's Republic of Bangladesh, the Department of Foreign Affairs, Trade and Development (DFATD), Canada, the Swedish International Development Cooperative Agency (SIDA) and the Department for International Development (UK Aid).  We gratefully acknowledge these donors for their support and commitment to icddr,b's research efforts. We also thank past programs at NOAA, also jointly with EPRI and NSF, for their support of research at the interface of infectious diseases and climate; these programs enabled the kinds of models presented here, and supported interdisciplinary efforts in a currently neglected area.

\newpage

\makeatletter
\renewcommand*\thefigure{S\arabic{figure}}
\renewcommand*\thetable{S\arabic{table}}
\makeatother
\setcounter{figure}{0}
\setcounter{table}{0}

\newpage
\section*{Supporting Information}
\subsection*{Process-based model and parameter estimation}

The population dynamics of cholera are represented by an SIRS (Susceptible-Infected-Recovered-Susceptible) model (equation 1) in which the population in subdivided into the following classes: $S$ for naive individuals who are susceptible to disease, $I$ for infected and infectious individuals, and $R$ for those who have recovered and have acquired immunity to the disease. This representation allows for temporary immunity with individuals in $R$ eventually returning to $S$ at a given rate $\phi$. The set of equations is given by:

\begin{align}
 \label{eq1}
 \begin{split}
&\frac{dS}{dt} = \Big(\mu P + \frac{dP}{dt} \Big) + \phi R - \lambda(t) S  - \mu S\\
&\frac{dI}{dt} = \lambda(t) S -\gamma I - \mu I \\
&\frac{dR}{dt} = \gamma I - \phi R - \mu R \\
\end{split}
\end{align}

where $P$ represents the population size, and $S$, $I$ and $R$, the number of individuals in those respective classes. The force of infection $\lambda$  denotes the rate of infection per individual susceptible in the population, which is given by: 
\begin{equation}
\lambda (t) = e^ { -\nu (t - t_0)} \bigg[\beta(t) \frac{I}{N} + \omega  \bigg]\\ \label{eq2}
\end{equation}

where the first term implements a long-term trend starting at the initial time $t_0$, and in the second term, $\omega$  represents a fixed background infection due to an environmental reservoir and $\beta$ refers to the transmission rate per infection: 
\begin{equation}
 \beta (t) = \exp \bigg[ \sum_{j=1}^6 b_{j} s_{j}(t)\; + \; \sum_{m=4}^5 b_{Em} s_{m}(t)\;  f(ENSO) \bigg] \; \bigg[ \frac {d\,\Gamma} {dt} \bigg] \\ \label{eq3}
\end{equation}

The expression for the transmission rate $\beta$ includes a seasonal component implemented through six splines $s_j$  with six coefficients $b_j$  determining the weight of each component, allowing for a flexible representation of the seasonality (Figure S1A). Climate forcing by ENSO is incorporated by considering the SST anomalies in the Ni\~no3.4 region of the subtropical Pacific for the month of January (http://www.cpc.ncep.noaa.gov/data/ indices/sstoi.indices). Specifically, this covariate is included in the terms relevant to the fall months (fourth and fifth splines), based on previous studies of the observed correlations between ENSO and cholera cases in this region \cite{pascual2000, bouma2001}. The transmission rate also includes a stochastic component (through a Gamma distribution $\Gamma$) to represent environmental or other sources of variation not accounted for by seasonality or ENSO. 

Following Reiner \textit{et al}. (2012), the El Ni\~no index is integrated into a sigmoidal function, where $ENSO_{Jan}$ refers to the anomaly of the ENSO reported for January at time $t$, normalized between -1 and 1 (Figure S1B). This functional form allows for nonlinear responses to ENSO and is given by the following expression: 
\begin{equation}
 f(ENSO) = A \bigg[\frac {\tan{(h\ ENSO_{Jan} (t))}} {\tan{(h)}} \bigg] \\ \label{eq4}
\end{equation}

We formally relate the model to data by assuming that only a fraction $\rho$ of new infections are detected by the surveillance methods ($\rho$ is estimated from data along with other model parameters, see below), and model the data as negative binomially distributed around these new infections, as follows. From equation 1, the rate of new infections is $\lambda S$, which gives the total new infections $\Delta I_k$ over a time interval $(t_{k-1}, t_k)$ as  $\Delta I_k = \rho \int_{t_{k-1}}^{t_k} \lambda S(t) dt $, and the measurement model that couples $\Delta I_k$ to the observed data $y_k$ at time $t_k$ is given as $y_k \sim NegBin(\Delta I_k, \sigma^2_{obs})$, where $NegBin(a,b)$ is the negative binomial distribution with mean $a$ and variance $a + a^2 b$, and $\sigma^2_{obs}$ denotes the variance parameter associated with detection uncertainty.

To evaluate the forecasting abilities of the model, we first fitted the model to the data between 1995-2010. Specifically, we carried out a likelihood-based inference via an iterated particle filtering method to estimate the parameters and initial conditions, to obtain the MLE (Maximum Likelihood Estimates) \cite{ionides2006, king2016, pomp}. Because we wish to reproduce as close as possible the conditions under which one would be generating the forecasts, we then updated the estimated parameters by extending the fit of the model to the `new' data that would have become available each year. This resulted in refining the fit of the model over a moving window of 12 months, starting from January 2011. The updated parameters and initial values of the state variables each January were used to simulate forward for the next year. 

To compare the predictions to the data for the years between 2011 and 2015, we first defined an outbreak as a season in which the total number of cases exceeds a given threshold. The threshold itself was determined based on the empirical distribution of the average number of cases between August and December for the period 1995-2010. Based on the predictions generated by 1000 simulations, we computed the predicted probability of surpassing the given threshold (top 50\%, 25\% and 5\% of all of the data). These respective probabilities can be interpreted as an outbreak, large outbreak and extreme outbreak respectively.\\ 

\subsection*{Figures}

\begin{figure*}[h!]
\centerline{\includegraphics[width=7in]{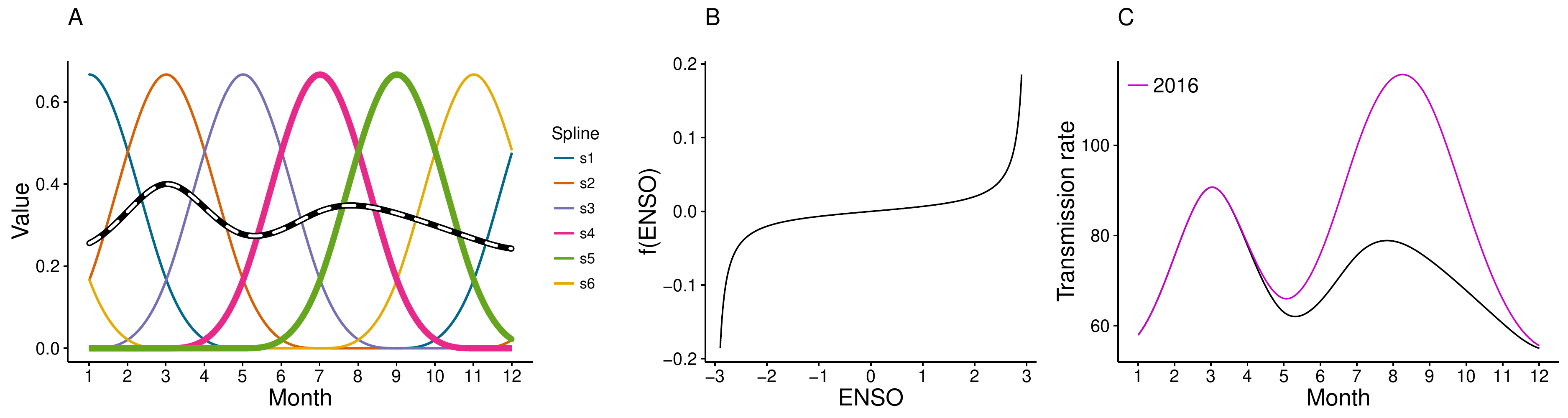}}
\caption{\textbf{A)} The six periodic splines considered in the seasonality component of the transmission rate (equation 3). The fourth and fifth splines were used to incorporate the effect of ENSO (thicker lines). \textbf{B)} Functional form of ENSO (equation 4), estimated with the parameters from the MLE. \textbf{C}) Transmission rate by month. The expression for 2016 is shown in magenta, while the black line illustrates a year with ENSO $\sim 0$. }
\label{fig:univariate_summary}
\end{figure*}

\newpage
\bibliography{choleraForecastDhaka_finalMS}

\begin{thebibliography}{10}

\bibitem{colwell1996}
Colwell RR.
\newblock Global climate and infectious disease: the cholera paradigm.
\newblock Science. 1996;274(5295):2025.

\bibitem{pascual2000}
Pascual M, Rod{\'o} X, Ellner SP, Colwell R, Bouma MJ.
\newblock Cholera dynamics and El Nino-southern oscillation.
\newblock Science. 2000;289(5485):1766--1769.

\bibitem{bouma2001}
Bouma MJ, Pascual M.
\newblock Seasonal and interannual cycles of endemic cholera in Bengal
  1891--1940 in relation to climate and geography.
\newblock In: The Ecology and Etiology of Newly Emerging Marine Diseases.
  Springer; 2001. p. 147--156.

\bibitem{rodo2002}
Rodo X, Pascual M, Fuchs G, Faruque A.
\newblock ENSO and cholera: a nonstationary link related to climate change?
\newblock Proceedings of the national Academy of Sciences.
  2002;99(20):12901--12906.

\bibitem{koelle2005}
Koelle K, Rod{\'o} X, Pascual M, Yunus M, Mostafa G.
\newblock Refractory periods and climate forcing in cholera dynamics.
\newblock Nature. 2005;436(7051):696--700.

\bibitem{pascual2008}
Pascual M, Chaves L, Cash B, Rod{\'o} X, Yunus M.
\newblock Predicting endemic cholera: the role of climate variability and
  disease dynamics.
\newblock Climate Research. 2008;36(2):131--140.

\bibitem{cash2008}
Cash BA, Rod{\'o} X, Kinter~III JL.
\newblock Links between tropical Pacific SST and cholera incidence in
  Bangladesh: role of the eastern and central tropical Pacific.
\newblock Journal of Climate. 2008;21(18):4647--4663.

\bibitem{matsuda2008}
Matsuda F, Ishimura S, Wagatsuma Y, Higashi T, Hayashi T, Faruque A, et~al.
\newblock Prediction of epidemic cholera due to Vibrio cholerae O1 in children
  younger than 10 years using climate data in Bangladesh.
\newblock Epidemiology and infection. 2008;136(01):73--79.

\bibitem{hashizume2008}
Hashizume M, Armstrong B, Hajat S, Wagatsuma Y, Faruque AS, Hayashi T, et~al.
\newblock The effect of rainfall on the incidence of cholera in Bangladesh.
\newblock Epidemiology. 2008;19(1):103--110.

\bibitem{akanda2009}
Akanda AS, Jutla AS, Islam S.
\newblock Dual peak cholera transmission in Bengal Delta: A hydroclimatological
  explanation.
\newblock Geophysical Research Letters. 2009;36(19).

\bibitem{hashizume2009}
Hashizume M, Wagatsuma Y, Faruque AS, Hayashi T, Armstrong B.
\newblock Climatic components of seasonal variation in cholera incidence.
\newblock Epidemiology. 2009;20(6):S153.

\bibitem{akanda2011}
Akanda AS, Jutla AS, Alam M, de~Magny GC, Siddique A, Sack RB, et~al.
\newblock Hydroclimatic influences on seasonal and spatial cholera transmission
  cycles: implications for public health intervention in the Bengal Delta.
\newblock Water Resources Research. 2011;47(3).

\bibitem{nasr2015}
Nasr-Azadani F, Unnikrishnan A, Akanda A, Islam S, Alam M, Huq A, et~al.
\newblock Downscaling river discharge to assess the effects of climate change
  on cholera outbreaks in the Bengal Delta.
\newblock Climate Research. 2015;64(3):257--274.

\bibitem{jutla2015}
Jutla A, Akanda A, Unnikrishnan A, Huq A, Colwell R.
\newblock Predictive Time Series Analysis Linking Bengal Cholera with
  Terrestrial Water Storage Measured from Gravity Recovery and Climate
  Experiment Sensors.
\newblock The American journal of tropical medicine and hygiene.
  2015;93(6):1179--1186.

\bibitem{reiner2012}
Reiner RC, King AA, Emch M, Yunus M, Faruque A, Pascual M.
\newblock Highly localized sensitivity to climate forcing drives endemic
  cholera in a megacity.
\newblock Proceedings of the National Academy of Sciences.
  2012;109(6):2033--2036.

\bibitem{martinez2016}
Martinez PP, King AA, Yunus M, Faruque ASG, Pascual M.
\newblock Differential and enhanced response to climate forcing in diarrheal
  disease due to rotavirus across a megacity of the developing world.
\newblock Proceedings of the National Academy of Sciences.
  2016;113(15):4092--4097.

\bibitem{cash2014}
Cash BA, Rod{\'o} X, Emch M, Yunus M, Faruque AS, Pascual M.
\newblock Cholera and shigellosis: Different epidemiology but similar responses
  to climate variability.
\newblock PloS one. 2014;9(9):e107223.

\bibitem{ionides2006}
Ionides E, Bret{\'o} C, King A.
\newblock Inference for nonlinear dynamical systems.
\newblock Proceedings of the National Academy of Sciences.
  2006;103(49):18438--18443.

\bibitem{king2016}
King A, D N, Ionides E.
\newblock Statistical Inference for Partially Observed Markov Processes via the
  R Package pomp.
\newblock Journal of Statistical Software. 2016;69(1):1--43.

\bibitem{roy2015}
Roy M, Bouma M, Dhiman RC, Pascual M.
\newblock Predictability of epidemic malaria under non-stationary conditions
  with process-based models combining epidemiological updates and climate
  variability.
\newblock Malaria journal. 2015;14(1):1.

\bibitem{king2015}
King AA, Domenech~de Cell{\`e}s M, Magpantay FMG, Rohani P.
\newblock Avoidable errors in the modelling of outbreaks of emerging pathogens,
  with special reference to Ebola.
\newblock Proceedings of the Royal Society of London B: Biological Sciences.
  2015;282(1806).

\bibitem{faisal2003}
Faisal I, Kabir M, Nishat A.
\newblock The disastrous flood of 1998 and long term mitigation strategies for
  Dhaka City.
\newblock In: Flood Problem and Management in South Asia. Springer; 2003. p.
  85--99.

\bibitem{perez2016}
Javier PS, King AA, Rinaldo A, Mohammad Y, G FAS, Mercedes P. Climate-driven
  endemic cholera is modulated by human mobility in a megacity; 2016.
\newblock In~review.

\bibitem{xie2007}
Xie P, Chen M, Yang S, Yatagai A, Hayasaka T, Fukushima Y, et~al.
\newblock A gauge-based analysis of daily precipitation over East Asia.
\newblock Journal of Hydrometeorology. 2007;8(3):607--626.

\bibitem{chen2008}
Chen M, Shi W, Xie P, Silva V, Kousky VE, Wayne~Higgins R, et~al.
\newblock Assessing objective techniques for gauge-based analyses of global
  daily precipitation.
\newblock Journal of Geophysical Research: Atmospheres. 2008;113(D4).

\bibitem{pomp}
King AA, Ionides EL, Bret\'o CM, Ellner SP, Ferrari MJ, Kendall BE, et~al..
  {pomp}: {S}tatistical Inference for Partially Observed {M}arkov Processes;
  2016.
\newblock R~package, version~1.5.
\newblock Available from: \url{http://kingaa.github.io/pomp}.

\end{thebibliography}

\end{document}